\documentclass[12pt,a4paper]{article}
\usepackage{a4wide,cite}
\usepackage[dvips]{epsfig}

\renewcommand{\thefootnote}{\fnsymbol{footnote}}

\begin{document}

 \thispagestyle{empty}
 \begin{flushright}
 {UTAS-PHYS-02-09} \\[3mm]
 {hep-th/0209233} \\[3mm]
 {July 2002}
 \end{flushright}
 \vspace*{2.0cm}
 \begin{center}
 {\Large \bf
   Three-body phase space: symmetrical 
   treatments\footnote{Presented by R.~Delbourgo 
   at the 15th Biennial Congress of the Australian Institute of Physics 
   (Sydney, Australia, July 2002)}}
 \end{center}
 \vspace{1cm}
 \begin{center}
 A.I.~Davydychev%
\footnote{
On leave from Institute for Nuclear Physics, Moscow State University,
119899 Moscow, Russia.
Email address: davyd@thep.physik.uni-mainz.de} \ \ 
    and \ \ 
 R.~Delbourgo%
\footnote{Email address: bob.delbourgo@utas.edu.au}
\\
 \vspace{1cm}
$^{a}${\em School of Mathematics and Physics,
University of Tasmania,\\[2mm] 
GPO Box 252-21, Hobart, Tasmania 7001, Australia}
\\
\end{center}
\hspace{3in}


\begin{abstract}
We derive expressions for three-body phase space that are 
explicitly symmetrical in the masses of the three particles, 
by three separate methods.
\end{abstract}


\newpage
\renewcommand{\thefootnote}{\arabic{footnote}}
\setcounter{footnote}{0}


\section*{Nature of the problem} 

The phase volume in $D$-dimensional space-time for the decay 
process $p\to 1+2+3$ can be reduced to the integral
\[
\rho_{p\rightarrow 1+2+3}=\frac{(4\pi)^{1-D}(p^2)^{1-D/2}}{2\Gamma(D-2)}\!
\int\!\!\int\!\!\int\!
{\rm d}s {\rm d}t {\rm d}u
\delta(s\!+\!t\!+\!u\!-\!m_1^2\!-\!m_2^2\!-\!m_3^2\!-\!p^2)
[\Phi(s,t,u)]^{D/2-2}\theta(\Phi),
\]
where $s=(p_1+p_2)^2$, $t=(p_2+p_3)^2$, $u=(p_3+p_1)^2$
are the three Mandelstam variables and, in the physical region, 
the Kibble cubic
\begin{eqnarray*}
\Phi(s,t,u)&=&stu - s(m_1^2m_2^2+p^2m_3^2) - t(m_2^2m_3^2+p^2m_1^2) -
 u(m_3^2m_1^2+p^2m_2^2) \nonumber \\
& & + 2(m_1^2m_2^2m_3^2 + p^2m_1^2m_2^2 + p^2m_2^2m_3^2 + p^2m_3^2m_1^2).
\end{eqnarray*}
stays positive. Clearly $\rho$ is symmetric in the three masses. 
However, when one eliminates one of the Mandelstam variables 
(or any linear combination) one is left, for even $D$, 
with an elliptic integral which is not {\em explicitly} symmetrical, 
although it must be so implicitly. The aim of this article is 
to exhibit three routes for getting a satisfyingly symmetrical 
result in the $4D$ case, where $\rho$ is nothing but the area 
of the Dalitz-Kibble plot~\cite{c1,c2} divided by $128\pi^3 p^2$. 
Detailed derivations are not given, for lack of space, 
but will be published elsewhere.

\section*{Route~1: Elliptic Drive}

The complicated expressions found by Almgren~\cite{c3} 
and Bauberger et al~\cite{c4}, obtained by finally integrating   
[$s_{1,2} = (m_1\pm m_2)^2$,  $s_{3,4}=(p\mp m_3)^2$ below]
\[
\rho = \frac{1}{128\pi^3 p^2}
\int_{s_2}^{s_3} \frac{{\rm d}s}{s} 
\sqrt{(s-s_1)(s-s_2)(s_3-s)(s_4-s)} \; ,
\]
involve the complete elliptic integrals,
$K(k)$, $E(k)$ and $\Pi(\alpha^2,k)$,
where 
\[
k=\sqrt{\frac{q_{+-}q_{-+}}{q_{++}q_{--}}}
\quad {\rm and} \quad
q_{\pm \pm} = (p\pm m_2)^2 - (m_1 \pm m_2)^2 \; .
\]
The main point is that the result is symmetric under 
$1\leftrightarrow 2$ interchange 
but {\em not obviously so} under $1\leftrightarrow 3$ and 
$2\leftrightarrow 3$ interchanges. 
The key to converting these expressions into symmetric form 
is to pass to the Jacobian zeta-functions $Z$ via 
\[
\frac{\Pi(\alpha^2,k)}{K(k)}
= 1 + \frac{\alpha Z(\beta, k)}
{\sqrt{(1-\alpha^2)(k^2-\alpha^2)}}
\]
with $\sin\beta = \alpha/k$ and to use the addition formula
(142.01) of Byrd and Friedman~\cite{c5}:
\begin{eqnarray*}
Z(\beta_1,k) \pm Z(\beta_2,k) = Z(\varphi_{\pm},k)
\pm k^2 \sin\beta_1 \sin\beta_2 \sin\varphi_{\pm} , 
\\
\tan\frac{\varphi_{\pm}}{2} = 
\frac{\sin\beta_1 \sqrt{1-k^2 \sin^2\beta_2}
      \pm \sin\beta_2 \sqrt{1-k^2 \sin^2\beta_1}}
      {\cos\beta_1 + \cos\beta_2} \; .
\end{eqnarray*}
After numerous manoeuvres, including identification 
of the three angles
\[
\varphi_2 = \varphi_{+}, \quad
\varphi_1 = \varphi_{-}, \quad
\varphi_3 =
\arctan\left[ \frac{\sqrt{q_{++} q_{--}}}{2(p m_3-m_1 m_2)} \right] \; , 
\]
etc. and noting the remarkable connection
\[
Z(\varphi_1,k) \!+\! Z(\varphi_2,k) \!+\! Z(\varphi_3,k)
= k^2 \sin\varphi_1\; \sin\varphi_2\; \sin\varphi_3 \; ,
\]
one obtains the beautifully symmetric form:
\begin{eqnarray*}
\rho&=&\frac{ \sqrt{q_{++} q_{--}}}{128\pi^3 p^2}
\Biggl\{(p^2+m_1^2+m_2^2+m_3^2) 
\left[ E(k)-K(k) \right]
\nonumber \\ && \hspace*{21mm}
+ \sqrt{q_{++} q_{--}} \; K(k) 
\Biggl[ \frac{Z(\varphi_1,k)}{\sin^2\varphi_1}
+ \frac{Z(\varphi_2,k)}{\sin^2\varphi_2}
+ \frac{Z(\varphi_3,k)}{\sin^2\varphi_3} \Biggl]
\Biggl\} .
\end{eqnarray*}

\section*{Route~2: Phase Cut Way}

In the triangular phase plot, pictured in Figure~1, 
points $O$ refer to configurations where the Mandelstam variables 
assume their minimum values, points $P$ where they attain 
their maximum values and the $Q$-lines and $R$-lines 
are tangents at $P$ and $O$.  
\begin{figure}[ht]
\begin{center}
\centerline{\vbox{\epsfysize=100mm \epsfbox{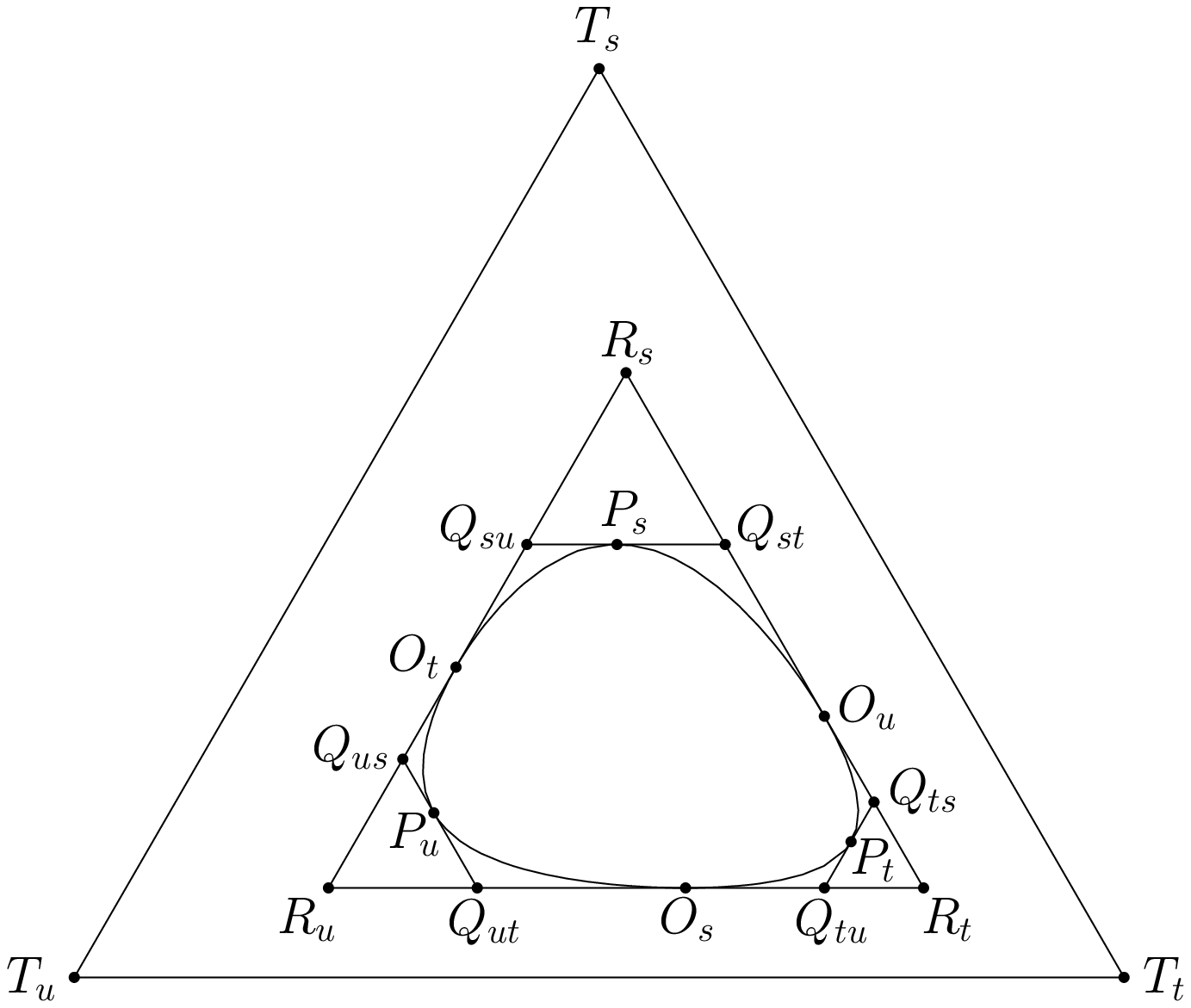}}}
\caption{Principal extremal points of the Dalitz--Kibble plot}
\end{center}
\end{figure}
Needing to integrate within the cubic outline $OPOPOP$, 
one is guaranteed to obtain a symmetric answer by finding 
the area of the hexagon $A_Q$ with $Q$ vertices and 
cutting out the six $OQP$ crescents. Now area
\[
A_Q = \frac{1}{2}\;(p-m_1-m_2-m_3)^2
\left[ (p+m_1+m_2+m_3)^2 - 4 (m_1^2+m_2^2+m_3^2)\right] \; ,
\] 
while a typical crescent has an area
\begin{eqnarray*}
A_{O_tP_sQ_{su}}
&=& \int_{s_{O_t}}^{(p-m_3)^2} \frac{{\rm d}s}{2s}
\Bigl[(p^2-m_3^2)(m_2^2-m_1^2)
-s \left( s-p^2-m_1^2+m_2^2+m_3^2+4m_2m_3 \right)
\\ && \hspace*{24mm}
- \sqrt{(s-s_1)(s-s_2)(s_3-s)(s_4-s)}
\Bigl]
\end{eqnarray*}
that is definitely given by an elliptic function. 
These crescents are readily evaluated in the limit 
of large $p$ and of small $Q =p-(m_1+m_2+m_3)$ 
by analytical means. One finds to leading order
\begin{eqnarray*}
A_{O_tP_sQ_{su}}& = &m_3^2 p^2 \left[
 \log\left(\frac{p}{m_2+m_3}\right) -\frac{3}{2}\right]
\\
 && -\frac{m_2m_3^2p^2}{12(m_2+m_3)^4}
 \left[12m_3^3+23m_2m_3^2+20m_2^2m_3+6m_2^3\right]
 + {\cal O}(p)\; ,
\\
A_{O_tP_sQ_{su}}&=&2Q^2 \left[ m_1m_2-\sqrt{m_1m_2m_3(m_1+m_2+m_3)}
     \;  \theta_{13}\right] + {\cal O}(Q^3)\; ,
\\
\sin\theta_{13}&=&\sqrt{\frac{m_1m_3}{(m_1+m_2)(m_2+m_3)}}\; , \quad
{\rm etc.}
\end{eqnarray*}
Subtracting all crescents and noting that 
 $\theta_{12}+\theta_{23}+\theta_{31}=\pi/2$
we get the full phase space area (in the two respective limits),
\[
A_{\Phi} = \frac{p^4}{2}-p^2 \left[ (m_1^2+m_2^2)
\log\left(\frac{p}{m_1+m_2}\right)+ {\rm permutations} \right]
+ {\cal O}(p)
\] 
or
\[
A_{\Phi} = 2\pi Q^2 \sqrt{m_1m_2m_3(m_1+m_2+m_3)} + {\cal O}(Q^3) \; .
\]
Higher order terms can be calculated quite systematically 
in this manner from the original $A_{OPQ}$ integrals for the crescents.

\section*{Route~3: Sunset Boulevard}

This method relies on the observation that~\cite{c6,c7,c8} 
phase space is twice the imaginary part of the sunset 
Feynman integral~\cite{c9}, possessing 3 internal lines, 
according to Cutkosky's cutting rules.  
Thus $\rho = 2 {\rm Im} I_3(p)$, where in $D$-dimensions,
\begin{eqnarray*}
I_3(p) &=& \int {\rm d}^D x \; \exp(-{\rm i} p\cdot x)\;
\prod\limits_{i=1}^3 \left[ {\rm i} \Delta_c(x|m_i) \right] \; ,
\\
{\rm i} \Delta_c(x|m_i) &=& \frac{1}{(2\pi)^{D/2}}
\left(\frac{m}{r}\right)^{D/2-1} K_{D/2-1}(mr) \; ,
\qquad
r=\sqrt{-x^2+{\rm i}\epsilon} \; .
\end{eqnarray*}
The Fourier integral leads to an integral of a Bessel 
function $J$ of the first kind with a product of three $K$ 
functions of the second kind; balanced progress may be achieved 
via the expansion, 
\[
K_{\nu}(z) = \sqrt{\frac{\pi}{2z}}
\left[ 1 + \frac{4\nu^2-1}{8z} + \ldots \right] \; ,
\]
which terminates for odd $D$ but is asymptotic for even $D$.
Since for example,
\[
K_1(m_1r) K_1(m_2r) K_1(m_3r) =
\frac{\pi^{3/2}\exp\left[-(m_1\!+\!m_2\!+\!m_3)r\right]}
{(8m_1m_2m_3 r^3)^{1/2}}
\left[ 1 + \frac{3}{8r}\left(\frac{1}{m_1} \!+\!
\frac{1}{m_2}\! +\! \frac{1}{m_3}\right) \!+\! \ldots \right]
\]
we are left with a doable integral, that produces 
a series of hypergeometric functions, whose leading term is 
\[
I_3(p) \sim \frac{1}{(\sigma \omega)^{(3-D)/2}}
\Gamma\left(\frac{3-D}{2}\right)\;
F\left( \frac{3-D}{4},\; \frac{5-D}{4}; \; \frac{D}{2}; 
\frac{p^2}{\sigma^2} \right) \; ,
\] 
with $\sigma=m_1+m_2+m_3$, $\omega=m_1m_2m_3$.
As imaginary parts of hypergeometric functions are~\cite{c8}
\[
{\rm Im} F(a,b;c;z) =
-\frac{\pi \Gamma(c) (z-1)^{c-a-b} \theta(z-1)}
{\Gamma(a)\;\Gamma(b)\Gamma(1-a-b+c)}\;
F(c-a,\; c-b;\; c-a-a+1;\; 1-z) \; ,
\]
we deduce that
\begin{eqnarray*}
&& \rho \propto \frac{(p^2-\sigma^2)^2}{\omega^{-1/2}\sigma^{7/2}}\;
F\left(\frac{9}{4},\; \frac{7}{4};\; 3;\; 
1-\frac{p^2}{\sigma^2}\right)
+{\cal O}\left( (p^2-\sigma^2)^3\right) \; ,
\\ &&
(p^2-\sigma^2) = Q (2m_1+2m_2+2m_3+Q) \; ,
\end{eqnarray*}
and recognize the threshold behaviour found by other routes. 
This is a particular case of the general $N$-body threshold 
behaviour, $\rho_N(p)\sim Q^{(N-1)D/2-(N+1)/2}$. 
One last comment: it is best to input the massless 
propagators ${\rm i}\Delta_c(x|0)=\Gamma(D/2-1)/(4\pi^{D/2}r^{D-2})$ 
at the stage of Fourier integration, rather than attempt 
to take the zero-mass limit of the above massive formulae, 
because that procedure is extremely tricky.

\section*{Acknowledgements}

We are pleased to acknowledge support from the Australian Research
Council under grant number A00000780.



\end{document}